\documentclass[usenatbib]{mnras}

\usepackage{newtxtext,newtxmath}

\usepackage[T1]{fontenc}

\DeclareRobustCommand{\VAN}[3]{#2}
\let\VANthebibliography\thebibliography
\def\thebibliography{\DeclareRobustCommand{\VAN}[3]{##3}\VANthebibliography}

\usepackage{graphicx}
\usepackage{subfigure}
\usepackage{xcolor}
\usepackage{amsmath,mathtools,bm}
\usepackage{bbm}
\usepackage{indentfirst}
\usepackage[english]{babel}
\usepackage{calc}
\usepackage{tikz}
\usepackage{hyperref}

\title[Is MOND an effect of cosmic frame-dragging à la Mach?]{Is MOND an effect of cosmic frame-dragging à la Mach?}

\author[Manuel Urueña Palomo]
        {Manuel Urueña Palomo$^{1}$\thanks{E-mail: yq87exim@studserv.uni-leipzig.de, muruep00@gmail.com}\\
$^{1}$Universität Leipzig, Faculty of Physics and Earth System Sciences, Linnéstraße 5, Leipzig (04103), Germany}

\pubyear{\the\year{}}

\begin{document}
\label{firstpage}
\pagerange{\pageref{firstpage}--\pageref{lastpage}}
\maketitle

\begin{abstract}
The physical origin of Modified Newtonian Dynamics (MOND), which predicts galaxy rotation curves via a single empirical acceleration constant $a_0$ without the need for dark matter, remains unknown. I propose a modified-inertia origin for both MOND's interpolating function and $a_0$ within a framework closely related to Gravitoelectromagnetism, the linearized limit of General Relativity. Following Sciama's model unifying gravity and inertia, which implements Mach's principle and the relativity of acceleration through gravitoelectric induction, I consider an idealized homogeneous universe undergoing constant Hubble expansion. I identify $a_0 \sim H_0^2R_u \sim c^2/R_u$ with the relative recession acceleration of cosmic matter near the horizon, which dominates the cosmological contribution. I show that this scale enters local inertial dynamics only when the relative acceleration of the universe becomes anisotropic and the Rindler horizon is comparable to or larger than the cosmic Hubble horizon, at accelerations of order $a_0$ and below. By the simplest Ansatz for the observer’s acceleration-dependent causal volume, accounting for the interplay between the cosmic and Rindler horizons, I derive MOND's simple interpolating function as a modified inertia law without the need to insert it by hand or make any further ad hoc adjustments. Physically, the region brought into causal connection ahead of the accelerating observer by the deformation of the cosmic horizon contributes less inertial drag than is lost from the region removed behind him by the Rindler horizon. This proof of concept motivates the possibility that MOND arises from large-scale linear inertial frame-dragging induced by the expanding Universe within General Relativity.
\end{abstract}


\section{Introduction}\label{sec1}

\subsection{Modified Newtonian Dynamics}\label{sec1.1}

The observed rotational velocities of stars and gas in galactic disks exceed the values expected from Newtonian gravity, as evidenced by flat galaxy rotation curves \citep{Rubin1983}. This discrepancy has motivated the development of modified theories of gravity and dynamics as alternatives to the still undetected physical dark matter hypothesis \citep{Review}. Mordehai Milgrom first identified the patterns in disk galaxies relating observed dynamics and visible matter only, such as the baryonic Tully–Fisher relation (the asymptotic rotational velocity scales as the fourth root of the baryonic mass), Renzo's rule (features in the visible mass distribution strongly correlate with features in the velocity profiles) and the empirical observation that Newton’s laws accurately predict dynamics only above a certain acceleration scale. Then, \cite{Milgrom1983a,Milgrom1983b,Milgrom1983c} proposed the simplest modification to Newtonian dynamics consistent with these observational constraints.

The pristine formulation of Milgrom's MOdified Newtonian Dynamics (or MilgrOmiaN Dynamics) \citep{WhitePaper,Banik2022,Famaey2012,Sanders2002} is a modification to Newton's second law of inertia or to Newton's law of gravity such that, at low accelerations or at the ‘deep-MOND' regime, the asymptotic rotational velocity becomes independent of the radial distance to the galactic core and satisfies $v^4=GMa_0$, with the need to introduce an empirically determined universal constant with dimensions of acceleration $a_0$. The transition between the classical Newtonian regime where $v^2=GM/r$ and the deep-MOND regime is governed by an ‘interpolating function' $\mu \left(a/a_0\right)$ such that, for Newton's second law,

\begin{equation}\label{eq1}
    \begin{split}
    \mathbf{F}_{\mathit{i,\, MOND}} = -m_i
    \mu\left(\frac{a}{a_0}\right)\mathbf{a}, \\ \mu\left(\frac{a}{a_0}\right) = \frac{a/a_0}{\left(1+\left(a/a_0\right)^n\right)^{1/n}},
\end{split}
\end{equation}

with $n=1$ for the simple and $n=2$ for the standard interpolating functions. With $a$ the Milgromian measured or ‘true' acceleration and the constant $a_0$ (also known as the Hubble acceleration or the acceleration scale constant), in the deep-MOND regime, $a\sim \sqrt{a_Na_0}$ with $a_N$ the usual Newtonian acceleration. The universal constant $a_0\sim 1.2 \times 10^{-10}$ m/s$^2$ defines the characteristic acceleration threshold below which MOND effects manifest. The interpolating function must satisfy the limiting conditions $\mu(x) \rightarrow x$ for $x\ll 1$ to reproduce the Tully-Fisher law, and $\mu(x) \rightarrow 1$ for $x\gg 1$ to recover the usual Newtonian laws at high accelerations. The asymptotic independence of the rotational velocity from distance reflects a scale-invariance property of the deep-MOND regime. Another requirement to match observations of non-isolated galactic systems is the external field effect (EFE), by which the external acceleration field $a_{e}$ cannot be ignored when estimating local dynamics, and can be approximated by $\mu \left(|(\mathbf{a_e}+\mathbf{a_i})|/a_0\right)\mathbf{a_i}$ with $\mathbf{a_i}$ the internal acceleration.

\newpage

In order to satisfy conservation laws and for computational purposes, Lagrangian-based models such as AQUAL \citep{AQUAL} and QUMOND \citep{QUMOND}, both special cases of three-potential TRIMOND theories \citep{TRIMOND}, have been developed. They recover Milgrom’s MOND in highly symmetric systems, while showing modest deviations elsewhere. Nevertheless, as non-relativistic models lacking a physical explanation for the origin of the interpolating function and the acceleration scale constant, and being insufficient to explain all dark matter related phenomena such as galaxy cluster dynamics, MOND is best regarded as an effective theory or an approximation to a more fundamental modification of gravity or inertia. Relativistic MOND theories, such as TeVeS from \cite{TEVES} and other modified gravity proposals with the same aim, such as MOG from \cite{MOG}, recover MOND's predictions for galaxies through extra scalar and vector fields at the expense of introducing several accompanying extra parameters and constants. MOND as modified inertia has been mainly explored modifying the kinetic action \citep{Milgrom2023,Milgrom2022,Milgrom2004,Milgrom1994}.

Progress in the foundations of physics often came along with the derivation of what were thought to be universal constants, such as Kepler's third law constant $K$ of each $i$ orbited body for a universal gravitational constant $K=4\pi^2/GM_i$. Notably, General Relativity (GR) was developed without the addition or removal of any fundamental constant, by borrowing the speed of light from Maxwell and the gravitational constant from Newton. Therefore, it is of considerable theoretical interest to explore MOND's single introduced acceleration scale constant as a derived parameter. \cite{Milgrom2020} has stressed since MOND's proposal that the coincidences of $a_0$ with cosmological parameters $a_0\sim cH_0 \sim c^2/R_u \sim c^2 \sqrt{\Lambda}$ may point to more basic theory underlying MOND phenomenology.

\subsection{MOND, Mach's principle, and cosmic expansion}\label{sec1.2}

The physical origin of MOND hinges on three main questions:

\begin{enumerate}

         \item[1)] whether it represents a modification to gravity or inertia,
         \item[2)] the nature of the characteristic acceleration scale constant $a_0$, and
         \item[3)] the precise form and derivation of the interpolating function $\mu(x)$,
  
\end{enumerate}

ordered from more general to more specific. Here I refer to the phenomenological origin and not to a complete FundaMOND theory. It is perhaps necessary to first identify why such a modification to gravity or inertia must be made and then proceed to its introduction in a complete relativistic theory. Once these three features are obtained as an approximation from an underlying basis requiring a modification of Newton’s laws, pristine MOND is entirely recovered. Featured descriptive effects such as the scale invariance and the approximated EFE follow from them.

Lagrangian-based MOND theories (e.g., AQUAL or QUMOND) modify gravity only, and introduce by hand both the interpolating function and the acceleration scale constant into the Lagrangian, failing to resolve 2) and 3). Modified inertia approaches do not settle 2), and relativistic MOND theories introduce even more constants and free parameters in order to obtain 2) and 3).

In contrast to these approaches, I take a different route and speculate on a Machian correspondence of 2) without assuming either possibility in 1) (which within a Machian framework, such a distinction might not even be necessary to be made). From that, I aim to find 3). Namely, I consider here the purely kinematical acceleration coincidence $a_0\sim cH_0$ within the context of Mach's principle and the relativity of acceleration, after our previous Machian attempt with the coincidence of a cosmic scalar $a_0\sim GM_u/R_u^2$ \citep{ManuelandSantander}.

The mechanism behind MOND can also be thought to come from two possibilities: A) it may arise from our current theories as we know them, such as from ignoring GR relativistic corrections at low accelerations, setting the wrong boundary conditions, or cosmic expansion interacting non-trivially; or B) it may indicate a foundational issue with GR, such as the postulate of the gravitational constant or the absence of both the relativity of acceleration and fulfillment of Mach's principle.

The A) possibility is akin to asking whether Newton's laws are always a good enough approximation to GR in the low velocities and shallow gravitational potentials of galaxies in our Universe. It is well known that GR reduces to Gravitoelectromagnetism (GEM) as a linear approximation in such limit retaining the effects of slow-moving mass currents, and GEM reduces to Newtonian gravity in the limit in which the gravitomagnetic field becomes negligible (in time-independent fields and negligible mass currents). Much literature has been built up against claims of GEM solving or significantly alleviating the galaxy rotation curve problem in recent years \citep{GEM1,GEM2,GEM3,GEM4}. A different case is whether ignoring cosmic expansion and assuming boundary conditions of asymptotic flatness may hide deviations from Newton's laws in galaxies, reviewed by \cite{Carrera2010}. Estimating the GEM contribution of McVittie's metric of a gravitational source mass embedded in a FLRW cosmological background, which accounts for the Hubble expansion of the Universe, \cite{Toth2021} found the gravitoelectric contribution to be negligible and the gravitomagnetic one to be vanishing. Thus, cosmological boundary conditions within standard GR seem not to account for galaxy rotation curves either. I therefore turn to B) and search for a fundamental realization missing in GR. 

There are substantial indications motivating a search for MOND in Mach's principle and the origin of inertia \citep{Barbour}. This connection was early noted by \cite{Milgrom1983a} and was later explored, though only to a limited extent, mainly in relation to the vacuum \citep{MilgromUnruh,Milgrom2006,Milgrom1999,Milgrom1994}. It has subsequently been pursued by \cite{ManuelandSantander,Corda2024,Corda2016,Anderson2011,Darabi2010,Gine2009}, as well as within another modified gravity proposal by \cite{MOGinertia}. \cite{Manuel} discussed in detail the broader aspects of Mach's principle that may be relevant to MOND. Only a brief set of these clues are developed in the following.
 
The most obvious connection between MOND and Mach's principle is the numerical coincidences of the acceleration scale constant $a_0$ with cosmological parameters. A widespread definition for Mach's principle is that local dynamics must be affected by the cosmic motion and mass distribution at large. Such a notion is only loosely satisfied by GR due to Birkhoff's theorem and the strong equivalence principle, even though GR is more Machian than classical Newtonian gravitation due to frame-dragging and the Lense-Thirring effect. One would then have reasons to think that if Machian effects not implemented in GR exist, they should become obvious when the matter in the Universe at large is more relevant than the local matter, i.e., in regions of shallow potentials or low field intensities. The latter is the key feature in MOND. The EFE of MOND also has a Machian reminiscence of such kind. Additionally, following Mach, the inertia of a body should decrease by having unaccelerated inertial masses removed from its vicinity (not satisfied by GR where inertia is absolute), and accelerated masses along with the body should not contribute to the increase of inertia. MOND can be thought of as modified inertia implementing a decrease of inertial masses at low accelerations, i.e., when far away from other masses.

I focus in this work on cosmic motion rather than only the mass distribution at large as done by \cite{ManuelandSantander}. The former would imply that MOND vanishes in a static universe, while the latter would necessitate an empty universe to do away with MOND.

MOND is fundamentally acceleration-based, and therefore requires a clear definition of acceleration itself: in particular, the frame with respect to which accelerations are to be measured. \cite{MilgromUnruh,Milgrom1999,Milgrom1994} claims that the vacuum provides such absolute frame by Unruh radiation. Others have explored Unruh radiation with the same purpose \citep{McCulloch2007}. Except for the case in which the vacuum would carry the information of the motion and distribution of matter at large, this is not purely Machian, since Mach's principle unequivocally gives meaning to accelerations relative to other masses. It is the mass and motion of the universe which define the absolute frame. Its motion is, on average and after smoothing out inhomogeneities, the one from the expanding universe. The question is therefore how particularly does the cosmic expansion of the matter distribution at large influence local galaxy dynamics?

This question was raised and explored by \cite{DarabiCosmic} in four proposals regarding flat rotation curves in an extensive although pertinent discussion for the present work. He first cleverly noticed that an acceleration imparted to a mass embedded in an accelerated expanding universe breaks the isotropy it had with respect to the radially accelerating masses in the universe. He assumed without further justification that a cosmic acceleration is induced on the mass in the same direction as the imparted acceleration in order to satisfy the equivalence principle. Such cosmic acceleration is assumed to be $a_c\sim cH_0$ to fit observations. He arrived at a minimum cosmic acceleration in dynamics and then argued that, to be compatible with the data, such extra acceleration had to be distance-dependent; vanishing both at the bulge of galaxies because of tight gravitational bounds and at very large distances from the bulge due to low accelerations recovering the isotropy of the cosmic accelerated expansion. I will expand on the initial interesting assessment of the breaking of isotropy.

\cite{Anderson2011} attempted to explain dark matter and dark energy starting from a Machian Heaviside-Sciama framework instead of Newtonian gravity, with an acceleration-dependent field falling as $1/r$. By assuming without further justification that the galaxy center constitutes the “local dynamic symmetry center” of the expanding universe (also invoked by \cite{Feoli2020} and \cite{Gupta2010}), an observer displaced from this center experiences an incomplete cancellation of the cosmological acceleration field, yielding an additional inward contribution. He then obtains a low angular velocity modification of effective inertia, after calibrating $\dot H$ to the desired galactic mass discrepancy, that does not result in the deep-MOND law or a constant $a_0$.

\cite{Gine2012} discussed MOND in relation to $a_0\sim cH_0$ as coming from the cosmic acceleration of distant matter. He considered that for a particle at the Hubble horizon, its equation of motion is $m_ia_0=F_e+GM_um_g/R_u^2$. But such a particle has a different observable universe than the one from the observer who sees the particle in his Hubble horizon, and there should be no force from $M_u$. \cite{Gine2009} also claimed that “the constant acceleration $a_0$ is the acceleration that feels an experimental body induced by the rest of the matter of the universe in its inertial frame of reference.”, even though $a_0=GM_u/R_u^2$ cannot be an acceleration from a gravitational force, since such force cancels out due to isotropy of the universe at large.

\subsection{Machian theories of inertia}\label{sec1.3}

There have been recurring attempts \citep{FayBraun2025} to unify gravity and inertia à la Mach by deriving the two in unison as early as the proposal of \cite{Reissner1915}, rediscovered by \cite{Schrodinger1925}, to the model of \cite{Treder1972}, including the well-known framework of \cite{Sciama1953} for the origin of inertia. These classical, linear, non-relativistic models fully implementing the relativity of accelerations share a Maxwellian framework similar to GEM, which is nothing but the linearized approximation to GR. But in contrast to GEM (and to GR), they offer a precise way to derive the value of the gravitational constant from physical parameters (and explain the weakness of the gravitational force), feature the equivalence principle as a consequence and not as an axiom, and explain inertia as of gravitational origin. I hereby use Treder's model \citep{Dennis2024} as an example, since it can be derived from a Lagrangian and avoids the anisotropy of inertial mass\footnote{It follows from considering relative velocities $\beta^2$ in the Lagrangian from Riemann's potential (and not the rate of change of the relative distances) that inertial mass is scalar, so that no anisotropy of inertia arises in such a model as suggested by \cite{Cocconi1958}. I thank Dennis Braun for recognizing such property of these relative velocity-dependent potentials.}, although the mechanism for the inertial force is equivalent to that of Sciama's model. 

Treder postulates a scalar and a vector potential for a matter density $\rho(r)$ in the form of

\begin{equation}\label{eq2}
    \begin{split}
    \varphi(\mathbf{r})=-\int \frac{\rho(\mathbf{r'})}{|\mathbf{r}-\mathbf{r'}|} d^{3}r',\\ \mathbf{A}(\mathbf{r})=-\int \frac{\rho(\mathbf{r'})\boldsymbol{\beta}(\mathbf{r'})}{|\mathbf{r}-\mathbf{r'}|} \, d^{3}r',
    \end{split}
\end{equation}

with $\boldsymbol{\beta}=\mathbf{v}/c$ where the source quantities are understood to be evaluated on the past light cone of $\mathbf r$. Therefore, after postulating that a freely moving body (which never experiences a force) follows a trajectory such that the total gravitational field generated by the rest of the universe vanishes $E_g=0$ in its instantaneous rest frame,

\begin{equation}\label{eq3}
    \begin{split}
    \mathbf{E_g}(\mathbf{r})=G\left(-\mathbf{\nabla}\varphi - \frac{2}{c} \frac{\partial\mathbf{A}}{\partial t}\right),
    \end{split}
\end{equation}

 giving the usual Newtonian gravitational acceleration and an extra term in the ‘gravitoelectric’ field being the inertial effect in terms of the acceleration of a body relative to the rest of the masses of the universe. Such force corresponds to an inertial drag with its origins in the gravitational relative motion of the mass of the universe

\begin{equation}\label{eq4}
    \begin{split}
    \mathbf{F}_{\mathit{i,\,Sciama}} = -m_iG \frac{2}{c} \, \frac{\partial \mathbf{A}}{\partial t}.
    \end{split}
\end{equation}

After some cancellations, (\ref{eq4}) gives the usual Newtonian inertial law $\mathbf{F}_{\mathit{i,\,Sciama}}=-m_ia$ for a finite universe of mass $M_u$ and radius $R_u$ provided that $G\equiv-c^2/2\varphi = c^2/(3M_u/R_u)$, which approximately holds in our Universe. In this way, local inertial forces result from the gravitational induction of the observable universe.

Note that since the gravitational constant $G\equiv-c^2/2\varphi$ is no longer constant, the volume from both integrals in $\varphi(\mathbf{r})$ and $\mathbf{A}(\mathbf{r})$ cancel out in the inertial term, and the inertial force (\ref{eq4}) does not explicitly depend on either the size or the total mass of the universe. Thus, adding an extra inhomogeneous mass density contributes to $-G\mathbf{\nabla}\varphi$ with an effectively fixed $G$ in the resulting equations of motion, and inertia remains Newtonian. This ensures that Treder's and Sciama's models reduce to the usual Newtonian inertial law, which has been an obstacle for relating such theories to modifications to inertia à la MOND, apart from the fact that MOND is a nonlinear theory.

The analogous effect is featured as linear frame-dragging in GR with a fixed gravitational constant $G$, whereby an accelerating mass distribution drags inertial frames by an acceleration

\begin{equation}\label{eq5}
    \begin{split}
    \delta\mathbf{a}({\mathbf{r}},t) = - \left(2+2\gamma\right)\frac{G\varphi(\mathbf{r},t)}{c^2}\mathbf{a},
    \end{split}
\end{equation}

in which $G\varphi(\mathbf{r},t)$ is the Newtonian gravitational potential of the mass distribution with acceleration $\mathbf{a}$, with $\gamma=1$ the post-Newtonian parameter of GR measuring how much spatial curvature is produced by unit rest mass (which takes an effective $\gamma=0$ in Sciama's and Treder's models, thus underpredicting GR's result by a factor of two) \citep{Nordtvedt}. Naturally, one requires again $-2G\varphi=c^2$ for complete dragging of inertial frames, $\delta\mathbf{a} =\mathbf{a}$, in the linearized limit and for inertia to be exclusively determined by the mass distribution of the universe à la Mach.

In his seminal 1983 paper introducing MOND, \cite{Milgrom1983a} stated that \textit{“Hopefully, \textbf{a theory can be found in which $\bm{a_0}$, which appears in the local dynamics and, for that matter, the function $\bm{\mu(x)}$ will be derivable from the theory and will turn out to depend on the distribution of mass in the universe (density) and its manner of expansion... }}[...]\textit{ If, for example, the inertia force is due to the interaction of the accelerated particle with an inertia field produced by totality of mass in the universe (see, for example, Sciama 1961) such that the inertia force is not proportional to the accelerations any more, the introduction of an acceleration constant into the local equations of dynamics is implied and $cH_0$ }[...]\textit{ is the natural cosmological acceleration parameter which can play this role.”}. I examine hereafter whether MOND can be derived from Sciama’s and Treder's models alone, despite previous attempts \citep{Roberts1984}, in an expanding universe.

\section{MOND from cosmic expansion and the relativity of acceleration}\label{sec2}

\subsection{The acceleration scale $a_0$ from cosmic dynamics}\label{sec2.1}

In an isotropically linear Hubble-expanding universe with uniform matter density, the mass enclosed within a sphere of radius $r$ grows proportionally to volume. Consequently, most of the observable and causally connected mass of the universe $M_u$ lies near the edge of the observable radius $R_u$ and appears to recede from the observer at a velocity $\sim c$, following Hubble's law $\mathbf{v}(r)=H_0r\mathbf{\hat{r}}$ where $H_0=100h$ (km/s)/Mpc is the present Hubble constant with $h\approx(0.67-0.73)$.

It is worth noting here that, for subluminal recession speeds, one cannot uniquely distinguish between motion through space and the expansion of space between source and observer, since the resulting kinematic redshift is observationally indistinguishable from cosmological redshift. As I am concerned about the gravitational and/or inertial effects that distant matter may have in local dynamics, I will assume that such effects are also indistinguishable in the same way. Thus, if we restrict ourselves to purely observable kinematics and how the distance between an observer and distant masses evolves as a function of time, following Hubble's law in a constant Hubble-expanding universe, the acceleration at a distance $r$ is given by the material derivative\footnote{A non-constant Hubble law would give $\mathbf{a}(r)=(\dot{H}+H^2)r\mathbf{\hat{r}}$. Only the recent past of our Universe experienced a positive ‘outward' acceleration $\mathbf{a}(r)$ after initial expansion, while deceleration dominated the distant past.} $\mathbf{a}(r)=D \mathbf{v}(r)/D t=H_0^2r\mathbf{\hat{r}}$. Most of the matter of the observable universe recedes relative to any ‘static' arbitrary observer with an $a\sim H_0^2R_u$, with an average of $a\sim (3/4)H_0^2R_u$ for the entire $M_u$, which is of the same cosmological order as $a_0\sim cH_0 \sim c^2/R_u$ up to factors depending on the precise cosmological radius used. I will refer to non-inertial observers in \textsection~\ref{sec2.2}. This result already suggests a way in which $a_0$ could appear in the Machian inertial force (\ref{eq4}). In such equation (and in induced inertial effects in GR, such as the Lense-Thirring effect), the induced inertial effect behaves as $1/r$, and the most distant cosmological regions still have the greatest effect on the inertial drag. Thus, the expected value of acceleration entering (\ref{eq4}) would be $\sim (2/3)H_0^2R_u$, as the $1/r$ gives relatively greater weight to nearby matter than an ordinary mass average does. If such were the origin of MOND's acceleration scale, this would entail that MONDian effects would vanish ($a_0 \rightarrow 0$) recovering Newton's laws if the expansion of the universe were to suddenly halt, given enough time for such an effect to reach the observer.

The above estimates serve only as an approximation, since a local theory must be sourced by signals reaching the observer at observation time. One cannot simply rely on present-day cosmological distances, nor can the calculation be done in terms of redshift or retarded proper distance: redshift depends on the source’s recession velocity at emission time, while retarded proper distances do not account for the increase in effective distance between the source and receiver caused by cosmic expansion. If distant masses were non-accelerating, then, within the GEM approximation, it might be reasonable to approximate the relevant sources by their values at current cosmological time. However, the radiative, transient part of the gravitational field differs from the constant-velocity contribution: its propagation is closer in spirit to distance traveled by light from emission to reception. Consequently, a more exact treatment would require integrating the relevant source terms over the past light cone in a FLRW spacetime, including the relativistic contributions to gravitation, such as pressure. The tension with cosmology expected from a Machian interpretation becomes most evident here. In searching for the origin of MOND as a theory removing the need for some or all dark matter, I nevertheless have to rely on a cosmological model whose matter content is dominated by dark matter and whose expansion history depends on that dark matter component. It is then convenient to use first a simplified toy model of a constant ($\dot{H}=0$) Hubble-expanding universe with uniform mass density after having addressed that dimensional estimations yield approximately the searched $a_0$ as the scale in which most visible mass appears to accelerate away from an inertial observer. This might serve as a good enough approximation to our recent Universe, given that a critical density FLRW universe is spatially flat, $\nabla\varphi=0$ at fixed cosmological time slices, and, as it will prove useful for the proposal of this work, because the causal effects of acceleration require some time to be experienced by the observer. The relative radial acceleration field of matter (only meaningful at one instant in time) for any static observer in such toy model can be conceptually described by

\begin{equation}\label{eq6}
    \begin{split}
    \frac{\partial \mathbf{v}}{\partial t}(r)=H_0^2 {r}\mathbf{\hat r}.
    \end{split}
\end{equation}

I have thus addressed question 2) on the nature of the acceleration scale constant $a_0$ as a first step towards the physical origin of MOND.

\subsection{Why $a_0$ only manifests at low accelerations}\label{sec2.2}

After identifying a possible origin for $a_0$ as the purely kinematical acceleration of mass distribution at large due to cosmic expansion, I turn to address question 3) regarding the origin of the interpolating function. 

\begin{figure}
\centering
\includegraphics[scale=0.36]{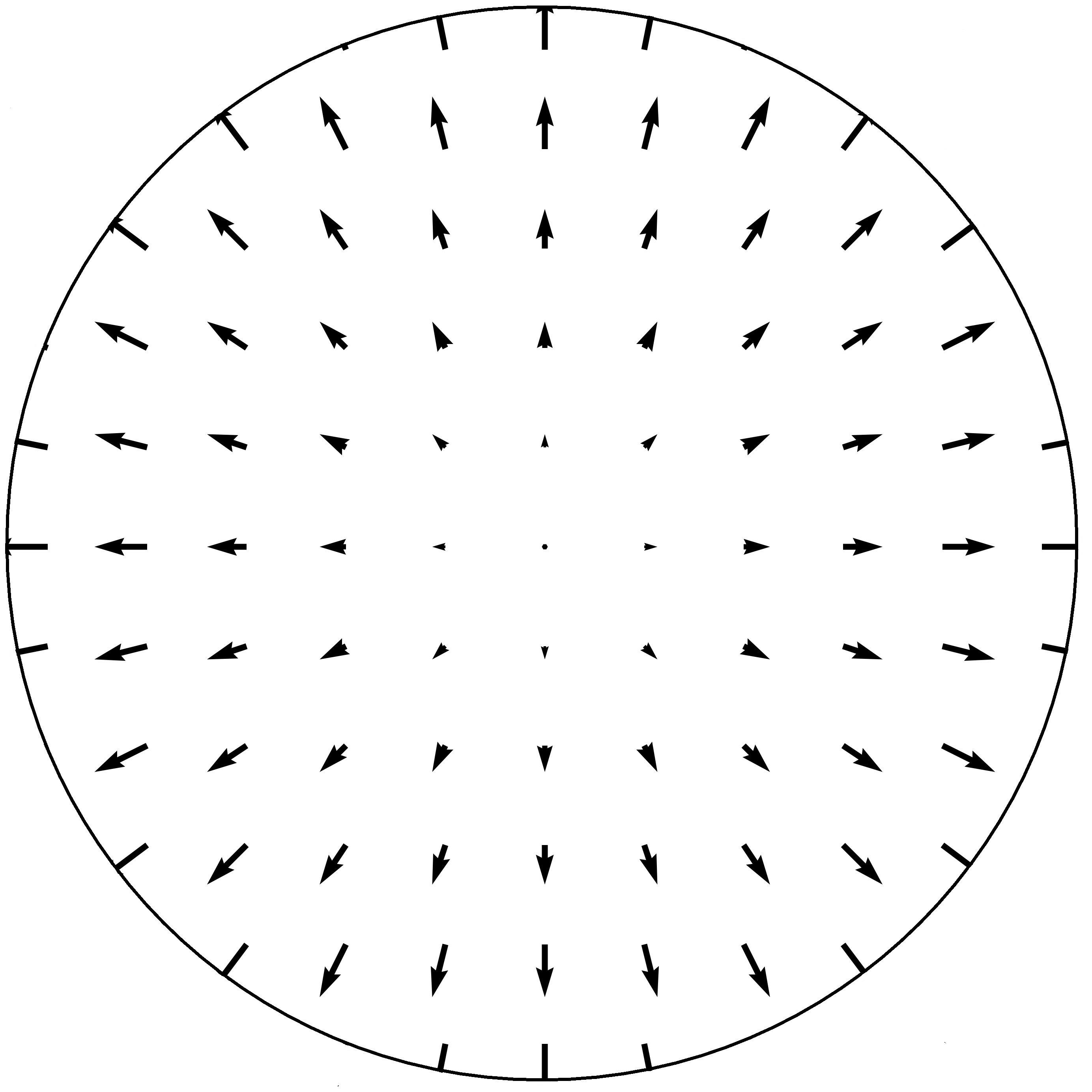}
\caption{A 2D slice of the acceleration field (\ref{eq6}) of a homogeneous, isotropic, constant Hubble expanding universe with constant matter density and a Hubble radius $R_u$, as perceived by any static centered observer. The field is isotropic and the most distant masses at $R_u$ recede at $H_0^2R_u\sim a_0$ from the inertial observer.}
\label{Fig1}
\end{figure}

The $a_0$ acceleration must enter local dynamics modifying Newton's laws at low accelerations of comparable magnitude, and Newtonian laws must be recovered for any higher acceleration. In order to identify the mechanism for such behavior of $a_0$, it will be convenient to picture the dynamics, in particular the acceleration field, of a uniform matter density due to cosmic expansion in the constant Hubble-expanding universe (\ref{eq6}) in Figure \ref{Fig1}.

\begin{figure}
\centering

\begin{minipage}{0.48\textwidth}
    \centering
    \includegraphics[scale=0.36]{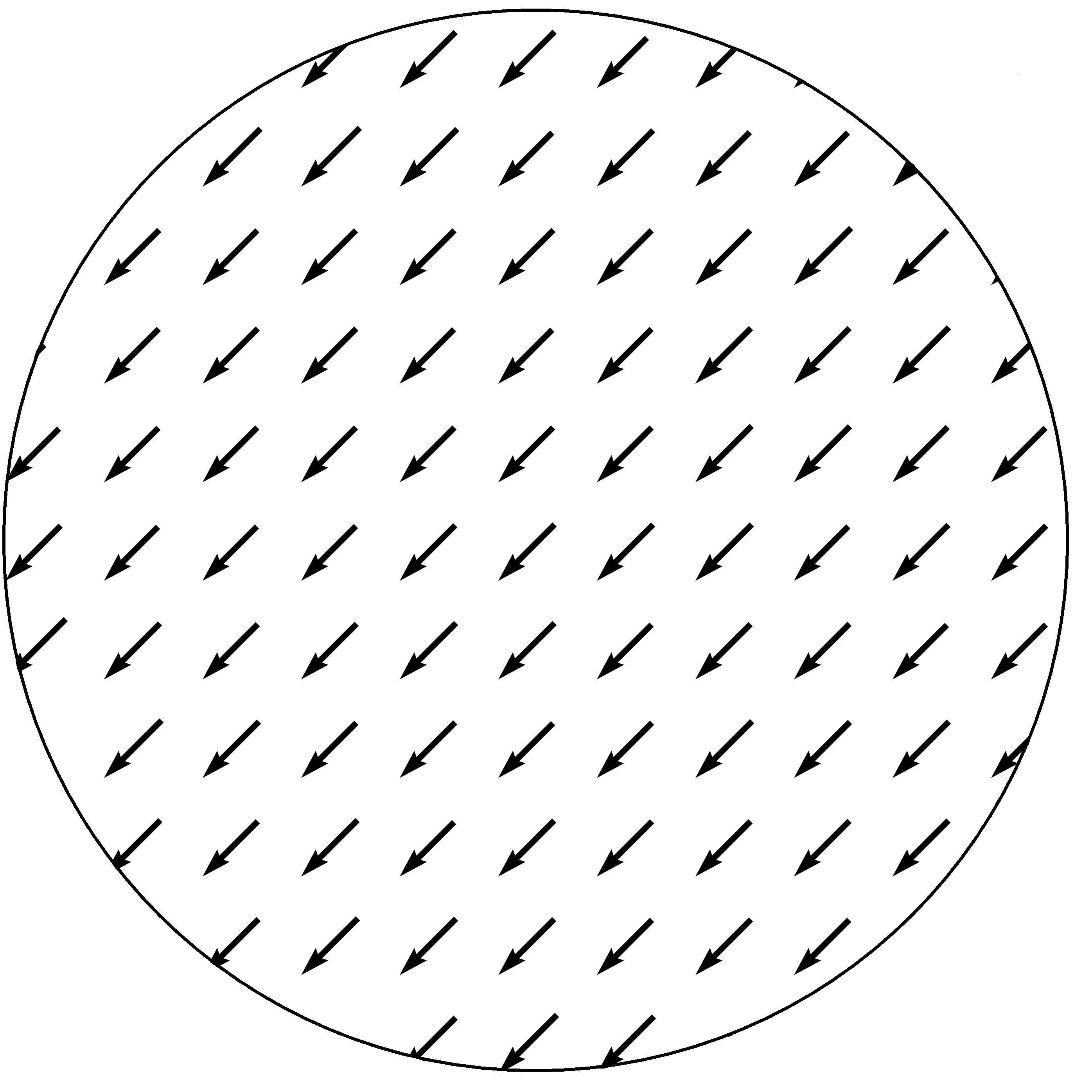}
    \caption{A 2D slice of the acceleration field (\ref{eq7}) of a homogeneous, isotropic, constant Hubble expanding universe with constant matter density and a Hubble radius $R_u$, as perceived by a centered accelerating observer to the top-right at $a\gg H_0^2R_u\sim a_0$. The same uniform-like field is perceived by an accelerating observer to the top right for any acceleration in a non-expanding universe with finite size $R_u$.}
    \label{Fig2}
\end{minipage}
\hfill
\begin{minipage}{0.48\textwidth}
    \centering
    \includegraphics[scale=0.36]{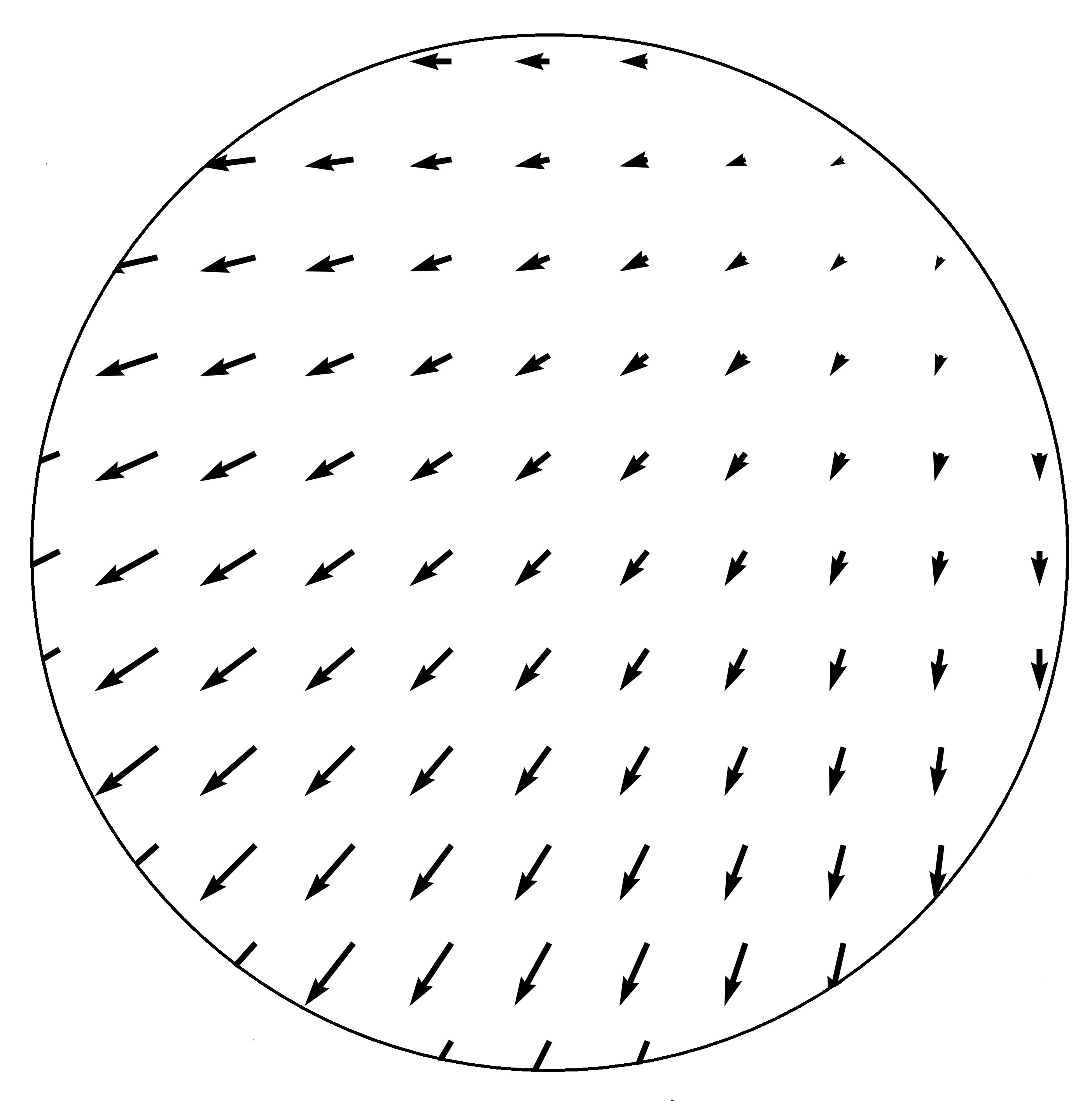}
    \caption{A 2D slice of the acceleration field (\ref{eq7}) of a homogeneous, isotropic, constant Hubble expanding universe with constant matter density and a Hubble radius $R_u$, as perceived by a centered accelerating observer to the top-right at $a\sim H_0^2R_u\sim a_0$.}
    \label{Fig3}
\end{minipage}

\end{figure}

If the observer undergoes a high acceleration $a\gg a_0$, and the principle of the relativity of accelerations holds, we may conceive the entire universe to be rigidly accelerating with respect to him in opposite direction, as depicted in Figure \ref{Fig2} for an acceleration of the observer to the top right $\mathbf{a}= a\mathbf{\hat a}$. There is practically no observational difference with an accelerating observer in a non-expanding static and finite universe.

But if the observer undergoes an acceleration as small as $a\sim a_0$, he perceives an anisotropic acceleration field of the distant matter, and not a uniform one. For instance, for an $a\sim H_0^2R_u$ acceleration to the top right, the perceived relative acceleration of the mass of the universe at large of a constant Hubble-expanding universe is Figure \ref{Fig3}. This anisotropy in the field still yields the usual Newtonian inertial law in Sciama's framework\footnote{Due to linear treatment of the vector potential and the superposition principle of both the uniform and the anisotropic fields, one giving zero inertial reaction and the other the usual inertial law.}. In the next section, I will argue how a MONDian law can arise after careful consideration of the causally connected volume of the universe.

For any acceleration of the observer with $\mathbf{a}=a\mathbf{\hat a}$, the relative acceleration field of the universe at large reads

\begin{equation}\label{eq7}
    \begin{split}
    \frac{\partial \mathbf{v}}{\partial t}(r,\theta,\phi)=H_0^2 {r}\mathbf{\hat r}- a\mathbf{\hat a}.
    \end{split}
\end{equation}

By the relativity of acceleration, I have shown why the acceleration scale $a_0$ appearing in MOND should be of the same order as the acceleration a body must have in order to experience MOND-like effects: if the accelerated motion of a body is such that large-scale mass distribution acquires a preferred acceleration scale relative to it, then this scale may appear in the law of inertia. In a non-expanding universe, there would be no reason for a modification to the laws of dynamics to happen at a particular acceleration scale. A possible mechanism for the EFE follows directly, as increasing $a$ pushes the observer back to the case of Figure \ref{Fig2} recovering an approximately uniform field for any $a\gg a_0$, although reproducing the detailed MOND EFE would require a complete nonlinear dynamical consideration.

\subsection{How $a_0$ modifies local dynamics}\label{sec2.3}

I have not yet addressed how $a_0$ would explicitly enter local dynamics, but only provided a reason why it may do so in an expanding universe at low accelerations, and not in a static one or at high accelerations. The acceleration scale of a body at which the isotropy of the relative acceleration field of the universe at large is broken is $\sim H_0^2R_u$, implying that such scale value must appear in the approximated inertial law (\ref{eq4}). While the requirement of a nonlinear relativistic generalization for the Machian inertia framework of Sciama and Treder might motivate enough a possible origin for MOND in a more fundamental theory, I continue to address question 3) and search the possible behavior of the interpolating function in the linear framework. Another property of MOND which is overlooked in this respect is that MOND's acceleration is always $a \ge a_N$, so whatever the effect of $a_0$ might be if having an origin in the cosmic expansion, it must always be a boost to local Newtonian acceleration with respect to the case of acceleration in a static universe. One could argue that if all the information from receding masses in the expanding universe is encoded in their gravitational mass $GM_u$, their distance to the body as a $1/r$ law, and their relative acceleration $\sim H_0^2r$, the combined $GM_uH_0^2r/r=GM_uH_0^2$ is independent of the distance in a suggestive MONDian way.

\cite{MilgromUnruh,Milgrom1999} identified that an accelerating observer perceives Unruh radiation, which could underlie MOND's interpretation as modified inertia. An additional relativistic effect takes place for such an observer: the Rindler horizon. It is well known that the appearance of a Rindler horizon at $R_R=c^2/a$ in a static universe causally disconnects part of the universe at large from the accelerating observer. In the presence of a cosmic horizon in a constant Hubble-expanding universe, the two horizons influence each other, thereby modifying the original causal structure. Even a small acceleration $a\ll H_0^2R_u$ biases the set of future signals intercepted by the observer: it enlarges the cosmic horizon in the forward direction, bringing into causal contact regions from which an inertial observer at the same initial position would not receive signals within any finite interval of their own proper time. This is not due to the peculiar velocity of the observer or length contraction, but to the observer ‘chasing' outgoing signals more effectively in the direction of acceleration, gaining causal access ahead.

I am concerned with Sciama's radiative term $\partial A/\partial t$ associated with propagating disturbances of the gravitational field, which is sensitive to the causal structure and to the observer-dependent causal boundary that determines which portions of the universe can contribute to the inertial drag. Given that acceleration does not change the past light cone of an event, and only alters where the observer will be by displacing its worldline, it merely instantaneously changes the received signals to those of a nearby event. It is only after a time $\sim c/a$, for which $a_0$ is several times the age of our Universe, that the observer can be considered to be causally disconnected from matter beyond the horizon. For the constant Hubble expanding universe I entertain here, only valid as an instantaneous representation of an expanding universe and serving only as an approximation for our recent past Universe, $R_u$ could be considered small enough to neglect the delay of the effect of the Rindler horizon on the observer. Thus, I assume onward that the delay in experiencing the effect from the Rindler horizon by the acceleration of the observer is negligible, and continue with the instantaneous description of such universe.

\begin{figure}
\centering

\begin{minipage}{0.48\textwidth}
    \centering
    \includegraphics[scale=0.36]{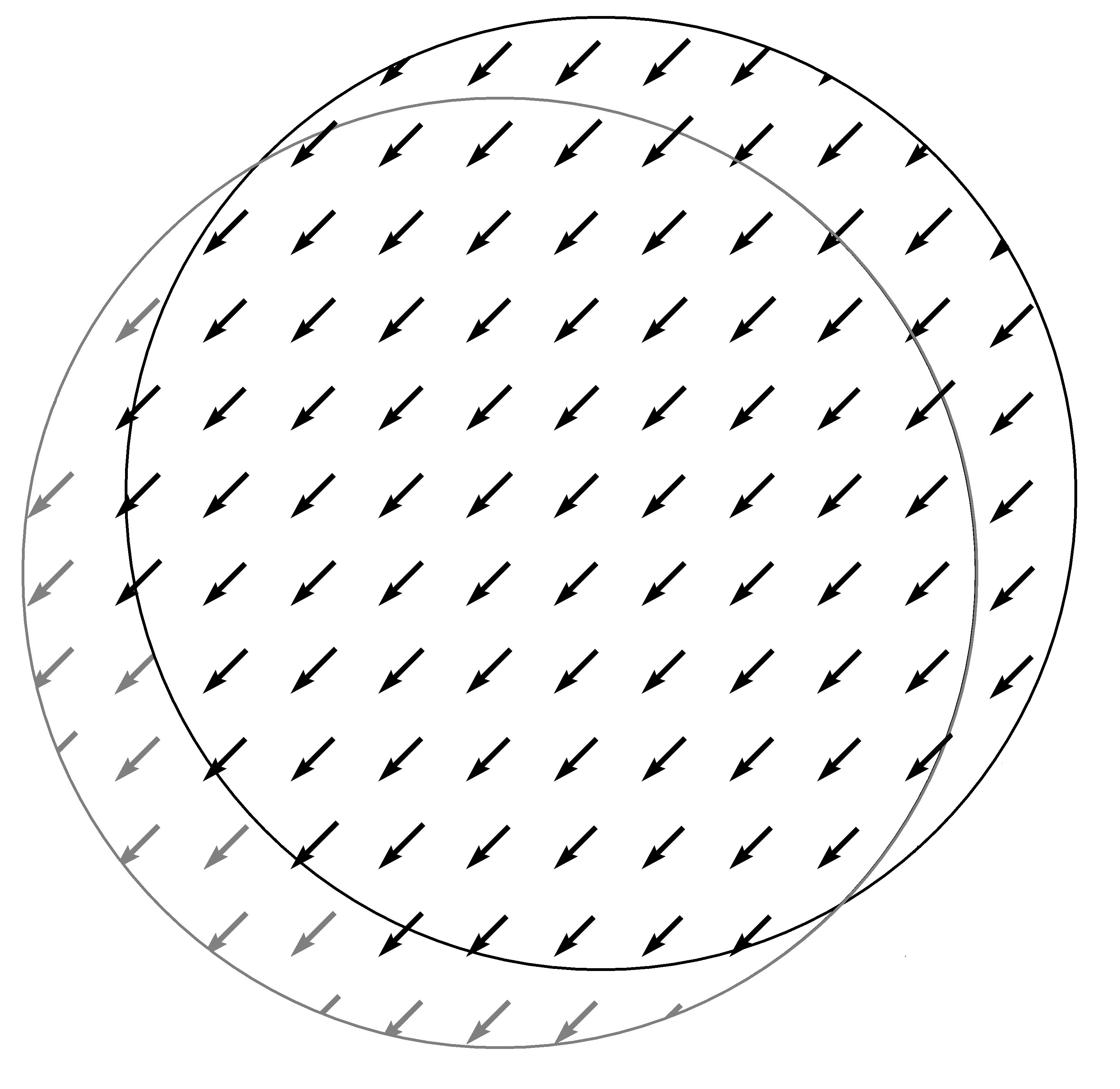}
    \caption{A 2D slice of the acceleration field (\ref{eq7}) of a homogeneous, isotropic, constant Hubble expanding universe with constant matter density and a Hubble radius $R_u$, as perceived by a centered accelerating observer to the top-right at $a\gg H_0^2R_u\sim a_0$. The black horizon (\ref{eq8}) accounts for the shifted observable volume with respect to the gray horizon of the static case due to the combined effect of the Rindler and cosmic Hubble horizon.}
    \label{Fig4}
\end{minipage}
\hfill
\begin{minipage}{0.48\textwidth}
    \centering
    \includegraphics[scale=0.36]{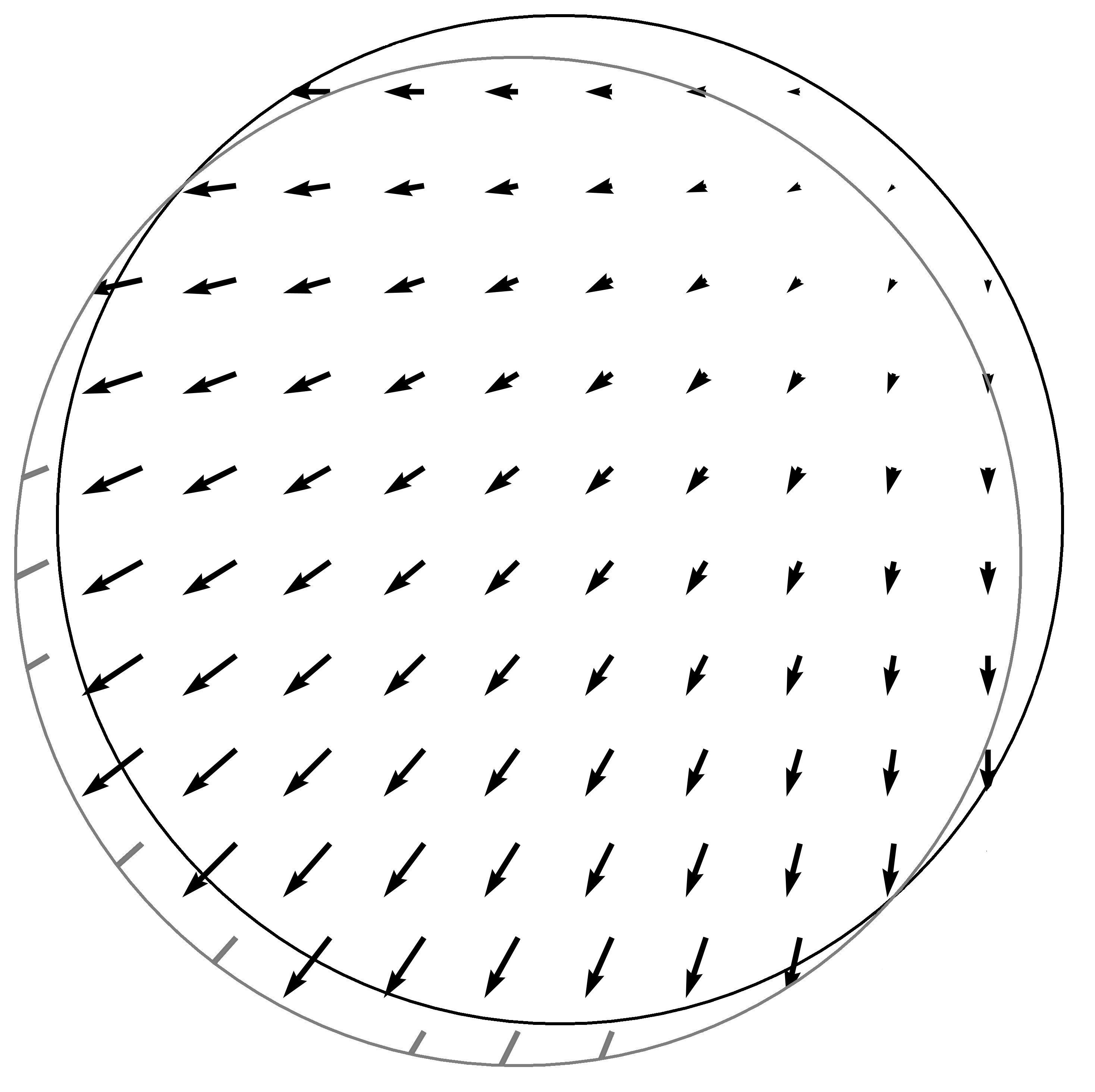}
    \caption{A 2D slice of the acceleration field (\ref{eq7}) of a homogeneous, isotropic, constant Hubble expanding universe with constant matter density and a Hubble radius $R_u$, as perceived by a centered accelerating observer to the top-right at $a\sim H_0^2R_u\sim a_0$. The black horizon (\ref{eq8}) accounts for the shifted observable volume with respect to the gray horizon of the static case due to the combined effect of the Rindler and cosmic Hubble horizon. Note that the new region ahead contributes less inertial drag than the lost one behind in gray.}
    \label{Fig5}
\end{minipage}

\end{figure}

For a static universe, and equally for a constant Hubble-expanding universe as long as $a\gg H_0^2R_u$, the causally ‘lost' uniform acceleration field of distant matter behind is compensated by a causally ‘gained' uniform acceleration field ahead (Figure \ref{Fig4}). Note here that since the Machian inertial force (\ref{eq4}) is independent of the volume or total mass integrated over, the resulting inertial drag is not altered by such change in causal structure, as long as the field remains uniform. This does not happen for an anisotropic field in the case of $a\sim H_0^2R_u$, where the ‘lost' part behind contributes much more to the inertial drag force than the ‘gained' part ahead due to the gradient in relative accelerations (Figure \ref{Fig5}). Hence, the horizon would no longer be a sphere centered on the observer but rather its shape must depend on the acceleration. By symmetry arguments, it is trivial that such modification to the inertial drag force does not change its direction. 

In order to estimate the acceleration-modified causal boundary, I consider for guidance a de Sitter space, where expansion is exponential with a constant Hubble parameter $H_0$ and there exists a cosmological horizon at $R_u=c/H_0$. The maximal symmetry of the de Sitter geometry provides a straightforward interpretation of the horizon associated with a uniformly accelerating observer. An accelerating observer maintaining a fixed spatial separation $R<R_u$ from an inertial one undergoes constant proper acceleration toward the inertial observer. Since the two observers can remain rigidly connected, the causal domain of the inertial observer forms part of the causal domain of the accelerating observer: any signal capable of reaching the inertial observer can also reach the accelerating one. Consequently, the horizon attributed to the accelerating observer may be interpreted geometrically as the spherical cosmological horizon of an inertial observer displaced in the direction of the acceleration $|\mathbf{r}-d(a)\mathbf{\hat{a}}|^2=R_u^2$. Moreover, the displacement of the sphere center $d$ must depend on acceleration through a dimensionless combination constructed from $a$, $R_u$ and $c$. Guided by this de Sitter analogy, I impose two minimal conditions for the causal boundary $r(\theta)$ of the accelerating observer. First, in the limit for $a\rightarrow\infty$, $d\rightarrow R_u$ and the causal boundary must approach $r(0)\rightarrow 2R_u$ in the forward direction of the acceleration, while it must collapse behind the accelerating observer $r(\pi)\rightarrow c^2/a$ as the Rindler horizon takes over. For any acceleration, the backward radial distance must satisfy $r(\pi)<c^{2}/a$. Second, in the inertial limit $a\rightarrow 0$, the usual causal boundary $r(\theta)\rightarrow R_u$ must be recovered. 

The simplest conceivable phenomenological Ansatz for such limits is a displaced sphere by $d(a)=R_u^2/(R_u+c^2/a)$ giving

\begin{equation}\label{eq8}
    \begin{split}
    r(\theta) = R_u \left[\frac{R_u}{R_u+\frac{c^2}{a}}\cos\theta + \sqrt{1-\left(\frac{R_u}{R_u+\frac{c^2}{a}}\right)^2 \sin^2\theta}   \right],
    \end{split}
\end{equation}

satisfying both conditions, and depicted in Figures \ref{Fig4} and \ref{Fig5}. 

Integrating for the Machian inertial drag force (\ref{eq4}) from the inductive term of (\ref{eq3}) over the shifted sphere volume (\ref{eq8}) with the acceleration field (\ref{eq7}) and applying superposition of fields results in

\begin{equation}\label{eq9}
    \begin{split}
    \mathbf{F}_{\mathit{i,\,Sciama}}=-m_i\frac{2}{c} \frac{G_0}{1-\frac{x^2}{3(1+x)^2}} \left. \frac{\partial \mathbf{A}}{\partial t} \right|_{\hat{\mathbf a}} - m_i\frac{2}{c} \frac{G_0}{1-\frac{x^2}{3(1+x)^2}} \left. \frac{\partial \mathbf{A}}{\partial t} \right|_{\hat{\mathbf r}},
    \end{split}
\end{equation}

with a fixed $G_0=c^2/(3M_u/R_u)$, accounting for the shifted sphere in $\varphi(a)$ by a factor in terms of $x\equiv R_u/R_R=R_u/(c^2/a)=a/(c^2/R_u)$. The integrals over the acceleration fields yield

\begin{equation}\label{eq10}
    \begin{split}
    \frac{\partial \mathbf{A}}{\partial t}\Bigr\rvert_{\mathbf{\hat{a}}} = \frac{3M_u a}{2cR_u} \left( 1-\frac{x^2}{3(1+x)^2} \right) \hat{\mathbf a},
    \end{split}
\end{equation}

and

\begin{equation}\label{eq11}
    \begin{split}
    \frac{\partial \mathbf{A}}{\partial t}\Bigr\rvert_{\mathbf{\hat{r}}} = - \frac{M_uH_0^2x}{c} \frac{(4x^2+10x+5)}{5(1+x)^3} \hat{\mathbf a},
    \end{split}
\end{equation}

resulting in the ‘$\rvert\mathbf{\hat{a}}$' field yielding the usual Newtonian inertia $m_ia$ due to being uniform, and the ‘$\rvert\mathbf{\hat{r}}$' field yielding an extra force term in opposite direction to the usual Newtonian inertia and proportional to $a$:

\begin{equation}\label{eq12}
    \begin{split}
    \mathbf{F}_{\mathit{i,\,Sciama}}=-m_i a \left[ 1 - \frac{2H_0^2R_u^2}{3c^2} \frac{4x^2+10x+5} {5(1+x)^3 \left( 1-\frac{x^2}{3(1+x)^2} \right)} \right] \hat{\mathbf a}.
    \end{split}
\end{equation}

The expression inside the brackets is already reminiscent of MOND as modified inertia, and it encodes the shape of the simple interpolating function, even though it does not yet render an interpolating function with the correct limits, as $[\eta(x=0)]\neq0$.

In addition to this inertial force, there is an extra gravitoelectric force $\mathbf{F}_{\varphi}=-m_i(-c^2/2\varphi(a))\mathbf{\nabla}\varphi(a)$ towards the center of the shifted sphere from the first term of the right hand side of the field (\ref{eq3}) giving

\begin{equation}\label{eq13}
    \begin{split}
    \mathbf{F}_{\varphi}=m_i\frac{G_0}{1-\frac{x^2}{3(1+x)^2}}\frac{M_u}{R_u^3}d(a)\mathbf{\hat{a}}=m_ia\frac{1+x}{2x^2+6x+3}\mathbf{\hat{a}},
    \end{split}
\end{equation}

for a homogeneous sphere volume whose center is displaced by $d(a)$, in the $+\mathbf{\hat{a}}$ direction, and further reducing inertial reaction.

The resulting Machian effective inertial force from (\ref{eq3}) reads

\begin{equation}
\label{eq14}
\begin{aligned}
\mathbf{F}_{\mathit{i,\,effective}}
&= \mathbf{F}_{\mathit{i,\,Sciama}} + \mathbf{F}_{\varphi} \\
&= -m_ia\left[1-\frac{2(4x^2+10x+5)}{5(1+x)(2x^2+6x+3)}-\frac{1+x}{2x^2+6x+3}\right]\hat{\mathbf a} \\
&\approx
-m_i a\left(\frac{x}{1+x}\right)\hat{\mathbf a},
\end{aligned}
\end{equation}

which is nothing but Milgrom's MOND inertial force given by the simple interpolating function $[\mu(x)]$ (\ref{eq1}) with $a_0=c^2/R_u$.

\section{Discussion}\label{sec3}

The origin of Modified Newtonian Dynamics (MOND) has long been loosely related to Mach's principle. Sciama's (and Treder's) well-known Machian model for unifying inertia and gravity implements the relativity of acceleration in a gravitational vector-potential theory of inertia which reduces to Gravitoelectromagnetism (GEM), i.e., to linearized General Relativity (GR), in a static finite universe when the total gravitational potential is of order $c^2$. I consider their inertial force induced by the relative acceleration of the universe at large for an accelerating body in a constant Hubble-expanding universe with uniform matter density. Most of the observable mass recedes kinematically relative to an inertial observer at an acceleration $H_0^2R_u \sim a_0$. Such acceleration scale is expected to enter local dynamics only when the relative acceleration of the universe for an accelerating body is anisotropic; when the acceleration of the body is $a\sim H_0^2R_u$ or lower, once the Rindler horizon is comparable to the cosmic Hubble horizon. I estimate the shape of the causally connected volume of the observable universe for an arbitrary acceleration of the body as a shifted sphere, analogous to the de Sitter spacetime case, which accounts for the interplay between both the Hubble cosmic horizon and the Rindler horizon. This shifted sphere horizon opens a causally connected region ahead of the accelerating observer, but this region contributes less inertial drag than the region removed behind by the Rindler horizon only at low accelerations. Integrating the Machian inertial gravitoelectric term over such volume and the relative acceleration field of the universe at large ($\ref{eq7}$), I derive the deep-MOND regime for low accelerations and the usual Newtonian inertia at high accelerations. The resulting Machian effective inertial force ($\ref{eq14}$) is nearly identical to MOND's simple interpolating function. It follows naturally that in the inhomogeneous universe case, e.g., by a galaxy next the observer, the gravitational pull is the usual Newtonian one and the effective inertial force is the one derived previously. In essence, at low accelerations, the universe appears to move in a manner that induces a weaker inertial reaction than at high accelerations, in which the universe appears to accelerate rigidly and the usual Newtonian inertia is recovered. The emergence of modified inertia in this way would suggest that neither the gravitational nor the particle action needs to be modified.

This derivation of MONDian dynamics addresses the main questions about the origin of MOND as modified inertia with $a_0$ arising from the relative acceleration of the constant Hubble-expanding matter density of the universe at large. Most remarkably, neither the interpolating function nor $a_0$ are introduced by hand. The function inside the shifted sphere horizon equation ($\ref{eq8}$) is chosen as the most elementary and natural function consistent with the required limiting conditions, and it specifies only the acceleration dependence of the causal geometry; it does not enter the final inertial force equation directly as an interpolating function. Regarding the scale $x=a/(c^2/R_u)$ which determines $a_0=c^2/R_u$, it is imposed by the limit behavior of the deformation of the accessible cosmic volume, which must depend on the ratio of the cosmic $R_u$ and Rindler $c^2/a$ horizons as $x=R_u/(c^2/a)$, with $R_u=c/H_0$. Sciama's inertial induction force ($\ref{eq12}$) with $K=2H_0^2R_u^2/3c^2$ contributes $K=2/3$ with $R_u=c/H_0$, while the gravitoelectric force ($\ref{eq13}$) provides the missing $1/3$ in the deep-MOND limit ($x\rightarrow 0$) in order to have $K=1$ and a well-behaved interpolating function\footnote{It was not acknowledged at the time of publication that the gravitoelectric force ($\ref{eq13}$) provides exactly the missing $1/3$, so $K=1$ was imposed by hand.}.

Regarding the external field effect (EFE), the acceleration determining whether a system lies in the MONDian or high-acceleration regime is its total acceleration relative to the cosmic inertial frame defined by the averaged large-scale mass distribution. For a subsystem with internal acceleration $\mathbf{a}_i$ embedded in a mother system producing an external acceleration $\mathbf{a}_e$, the total acceleration relative to the universe is $a=|\mathbf{a}_i+\mathbf{a}_e|$. Since the inertial response is nonlinear in this $a$, $\mathbf{a}_e$ cannot be removed from the internal equations of motion. To put it simply, $r(a)$ is unique and cannot be broken down into a linear sum of horizons. It becomes evident why one cannot study a dynamical system as isolated from others in MOND, since this ignores the universe at large which is responsible for the modification to the usual Newtonian laws. Regarding the internal field effect, whereby large internal accelerations within a composite system do not suppress the MONDian motion of its low-acceleration center-of-mass, I cannot find an explanation within my proposal other than taking $r(a_{\rm COM})$ for the center-of-mass acceleration, and hoping that this issue is naturally resolved in a more complete field theory in which inertial and gravitational effects share a common formal origin, as in GR.

With $R_u=c/H_0$ of the constant-Hubble model, it follows that $a_0=c^2/R_u=cH_0$ has the usual dependence on the radius of the observable universe $R_u$ (the larger the size of the observable universe, the more Newtonian it becomes) governed by the Hubble parameter $H_0$ (the stronger the expansion of the universe, the more MONDian it becomes). The $a_0=c^2/R_u$ motivates the ongoing discussion of the variation of $a_0$ at high redshifts. In the limit in which there is no cosmic expansion and $H_0\rightarrow 0$, there is no MONDian dynamics, aside from the usual problems of a static infinite universe.

If MOND were to have an origin in Sciama's inertial drag force and gravitoelectric induction in the GEM approximation, which corresponds to linear frame-dragging in GR, it could in principle also contribute to gravitational lensing. This would be particularly relevant given the excess lensing observed in galaxies attributed to dark matter, since frame-dragging modifies null geodesics in GR.

Given that a nonlinear inertial term emerges from a linear theory, one may ask whether other features also commonly regarded as necessary for a FundaMOND theory, such as violations of the strong equivalence principle and of Birkhoff’s theorem, are required by the equations governing MONDian dynamics. 

It became a prevailing view after \cite{Sciama1953} that only some cosmological solutions in GR were to be considered Machian, featuring a complete dragging of inertial frames, such as the isotropic FLRW models with critical density, in which $c^2 \sim -G\varphi$. These conditions allow one to interpret inertial reaction forces as of gravitational origin in the linear order approximation, as Sciama suggests. In his relativistic nonlinear extension of a generally covariant integral formulation of GR, \cite{Sciama1969} expressed the local metric determining the equivalent inertial forces in terms of a volume and a surface integral of the contributions from the stress-energy tensor and a retarded Green function within a specified horizon, with the surface integral representing the contribution to the metric from sources beyond the horizon. The vanishing of the surface integral implies inertial forces arising only by matter inside the horizon, satisfying Mach's principle, which occurs in all nonempty relativistic FLRW models. The evaluation of the volume integral at low accelerations remains a subject of ongoing research, which might result in the true interpolating function in agreement with observational constraints. Naively introducing the factor of $2$ missing in the linearized approximation to ($\ref{eq5}$), and rescaling $G$, leaves the shape of the interpolating function unchanged. I therefore conjecture that the true function arises primarily from the adopted cosmological model than from the nonlinear completion of the field equations themselves.

\section*{Acknowledgments}

I thank Dennis Braun from Bonn for translating and extending Treder's work, as well as for discussions of the inertial drag force of such models. I thank Harry Desmond for useful comments for v2.

\bibliographystyle{mnras}
\bibliography{bibliography}

\label{lastpage}
\end{document}